\documentclass{appolb}
\usepackage{graphicx}

\usepackage{amsmath}
\usepackage{hyperref}

\begin{document}

\title{Generalised Bohr Hamiltonian for Gogny interactions%
\thanks{Presented at the XXXVIII Mazurian Lakes Conference on Physics, Piaski, Poland, August 31 --September 6, 2025.}
}

\author{C.~Azam$^a$, D.~Davesne$^b$, Y.~Lallouet$^c$, L. Pr\'{o}chniak$^d$, M.~Frosini$^a$, A.~Pastore$^a$
\address{$^a$ CEA, DES, IRESNE, DER, SPRC, F-13108 Saint Paul Lez Durance, France\\
$^b$ Institut de Physique des 2 Infinis de Lyon, CNRS-IN2P3, UMR 5822, 
             43 Bd. du 11 Novembre 1918, F-69622 Villeurbanne cedex, Universit\'e Lyon 1, France\\
$^c$Lycée Malherbe, 14, Avenue Albert Sorel, 14000 Caen, France\\
$^d$ Heavy Ion Laboratory, University of Warsaw, PL-02-093 Warsaw, Poland
}
}

\maketitle
 \begin{abstract}
 Using a generalised Bohr Hamiltonian formalism together with the Gogny interaction to provide a microscopic input of the required mass parameters, we present the potential energy surface and the evolution of the first excited 2$^+$ state for the gadolinium isotopic chain. 
 We observe that the energies and electromagnetic transitions of low-lying states are fairly similar for the various parametrisations and are in good agreement with experimental data.
 \end{abstract}
  
 \section{Introduction}

 Mean field (MF) methods based on effective nucleon-nucleon (NN) interactions can now achieve a remarkable level of accuracy when describing nuclear ground state properties \cite{Goriely_2009}.
 The main advantage of using this approach compared to more fine-tuned macroscopic models is that the same NN interaction can be used to calculate excited states in conjunction with other many-body methods, such as the quasi-particle random phase approximation (QRPA), the rigid rotor method, and the generator coordinate method (GCM) \cite{ring2004nuclear}.
 Each of these many-body methods has its own advantages and disadvantages, depending on the observable in question. For instance, QRPA is better suited to describing the structure of giant resonances, whereas the rigid rotor is better suited to describing a rotational  spectrum~\cite{peru2014mean}.
Ideally, the GCM would represent a very useful tool to describe the nuclear spectrum also outside these limiting cases, but 
 depending on the number of collective coordinates employed, the computational cost of a GCM calculation can be quite high \cite{egi16,rob18}. Consequently, this method is not suitable for systematic calculations, particularly when combined with finite-range effective NN interactions.
 One way to overcome this problem without significantly compromising the description of the nuclear spectrum is to use the generalised Bohr Hamiltonian (BH). Although the BH model has been widely studied in connection with analytical potentials \cite{Fortunato_2005}, there has been recent renewed interest in using the BH model with potentials obtained from microscopic many-body (MF) models.
 Such an approach has been developed in Refs.\cite{kumar1967complete,Libert_1999,Prochniak_1999} and then applied systematically in Refs.\cite{bertsch2007systematics,del10,delaroche2024investigations} in order to analyse the properties of the lowest excited states of all even-even nuclei.
In this article, we carry out a series of calculations of the nuclear spectrum of the gadolinium isotopic chain using the BH and a selection of Gogny interactions, to check how sensitive the results are to the various parameterisations.

 \section{Formalism \label{sect:2}}

 In the original Bohr model \cite{Bohr_1952}, the collective excitations are obtained through the dynamical deformation of the nuclear surface. Considering only quadrupole degrees of freedom and working within the intrinsic frame of the nucleus, the relevant parameters are two collective coordinates $\beta,\gamma$ and three Euler angles. We refer to Ref.\cite{Nik09} for more details. The $\beta,\gamma$ coordinates are related to the expectation values of the quadrupole operators $\langle \hat{Q}_{2\mu}\rangle$ (in Cartesian coordinates)
 \begin{eqnarray}
   \hat{Q}_{20}=\sum_i(2z_i^2-x_i^2-y_i^2)\;\;, \;\;  \hat{Q}_{22}=\sum_i\sqrt{3}(x_i^2-y_i^2)\,,
 \end{eqnarray}
 via the relations
 \begin{eqnarray}
     \beta \cos\gamma = \sqrt{\frac{\pi}{5}} \frac{1}{A \langle r^2 \rangle }\langle \hat{Q}_{20}\rangle\;\;, \;\; \beta \sin\gamma = \sqrt{\frac{\pi}{5}} \frac{1}{A \langle{r}^2 \rangle }\langle \hat{Q}_{22}\rangle .
 \end{eqnarray}

 \noindent $A$ is the number of nucleons  and $\langle{r}^2 \rangle $ is the estimated value of the nuclear radius as extracted from the liquid drop model \cite{Libert_1999}.
 In the intrinsic frame, the collective Hamiltonian  has a simple form $ \hat{H}_\text{coll} = \hat{T}_\text{vib} + \hat{T}_\text{rot} + \hat{V}_\text{coll} (\beta,\gamma) $ where 
 $\hat{V}_\text{coll}$ is the collective potential energy, while the vibrational and rotational parts read respectively
 \begin{eqnarray}
     \hat{T}_\text{vib} & =&  -\frac{\hbar^2}{2\sqrt{G}} \left[ \frac{1}{\beta^4} \frac{\partial }{\partial \beta} \beta^4 \sqrt{G} \frac{B_{\gamma\gamma}}{G_\text{vib} } \frac{\partial }{\partial \beta} + \frac{1}{\beta^2 \sin 3\gamma}\frac{\partial }{\partial \gamma} \sin 3 \gamma \sqrt{G} \frac{B_{\beta\beta}}{G_\text{vib} } \frac{\partial }{\partial \gamma} \right. \nonumber \\
     && \left. - \frac{1}{\beta^4} \frac{\partial }{\partial \beta}  \beta^3 \sqrt{G} \frac{B_{\beta\gamma}}{G_\text{vib} } \frac{\partial }{\partial \gamma}  - \frac{1}{\beta \sin 3 \gamma}\frac{\partial }{\partial \gamma} \sin 3 \gamma \sqrt{G} \frac{B_ {\beta\gamma}}{G_\text{vib} } \frac{\partial }{\partial \beta} \right] \label{kin:vib},\\
     \hat{T}_\text{rot}  & =& \frac{1}{2} \left[\frac{\hat{L}_x^2}{\mathcal{I}_x}  +   \frac{\hat{L}_y^2}{\mathcal{I}_y}  +   \frac{\hat{L}_z^2}{\mathcal{I}_z} \right], \label{kin:rot}
 \end{eqnarray}
 with $G= G_{\text{vib}} B_x B_y B_z $ and $G_{\text{vib}} =  B_{\beta\beta} B_{\gamma\gamma} - B_{\beta\gamma}^2 $.  $\hat{L}_i$ denotes the intrinsic component of the total angular momentum and  $B_{\beta\beta}(\beta,\gamma)$, $B_{\beta\gamma}(\beta,\gamma)$, $B_{\gamma\gamma}(\beta,\gamma)$ are the vibrational mass parameters. Finally, we define the nuclear moments of inertia as
 \begin{equation}
     \mathcal{I}_\kappa(\beta,\gamma)= 4B_\kappa(\beta,\gamma)\beta^2 \sin^2\left(\gamma_\kappa \right) \quad \kappa=\{x,y,z\}.
 \end{equation}
\noindent where $\gamma_x = \gamma - \frac{2\pi}{3}$  $\gamma_y = \gamma + \frac{2\pi}{3}$ and  $\gamma_z = \gamma$. 
 In order to find the eigenvalues of the Hamiltonian, we use the methodology presented in  Refs. \cite{Libert_1999,Prochniak_1999,prochniak2009quadrupole} that consists in expanding $\hat{H}_{coll}$ on a suitable basis in the $\beta,\gamma$ plane and then diagonalising it. A new numerical code has been thus developed for this purpose and tested against analytic potentials and published results.

 \section{Results}

 The generalised BH uses the six mass parameters appearing in the kinetic terms given in Eqs.\eqref{kin:vib}-\eqref{kin:rot}, plus the collective potential. Rather than using analytical expressions, we derive them through constrained Hartree Fock Bogoliubov (HFB) calculations \cite{ring2004nuclear} based on the Gogny effective interactions:  D1S~\cite{decharge1980hartree}, D1N \cite{Chappert_2008}, D1M \cite{ Goriely_2009} and D3G3M \cite{BAtail_2025}. See Ref.\cite{frosini2021ab} for the numerical details of the adopted HFB solver.
 Of particular interest, the Gogny D1M and D3G3M have been adjusted including the zero-point energy \cite{girod1979zero} arising from the coupling to collective degrees of freedom.
 Such a quantity is typically added to the BH calculations \cite{Nik09} and may lead to slight inconsistency compared to the way the parametrisations are adjusted. This may be the case for shape-coexistence nuclei, but more systematic analysis are needed before drawing strong conclusions.
In this paper, the mass parameters are calculated using the perturbative cranking approximation \cite{baran2011quadrupole}. As a consequence, we perform a global scaling  by a factor of 1.3 as discussed in Ref.\cite{Libert_1999}. We recall that in recent years, some remarkable progress have been made to improve on this approximation ~ \cite{washiyama2024five}, but the method has not been applied systematically yet and more work is thus needed in order to assess systematic improvements. 

 \begin{figure}
     \centering
      \centering
         \includegraphics[width=0.5\textwidth]{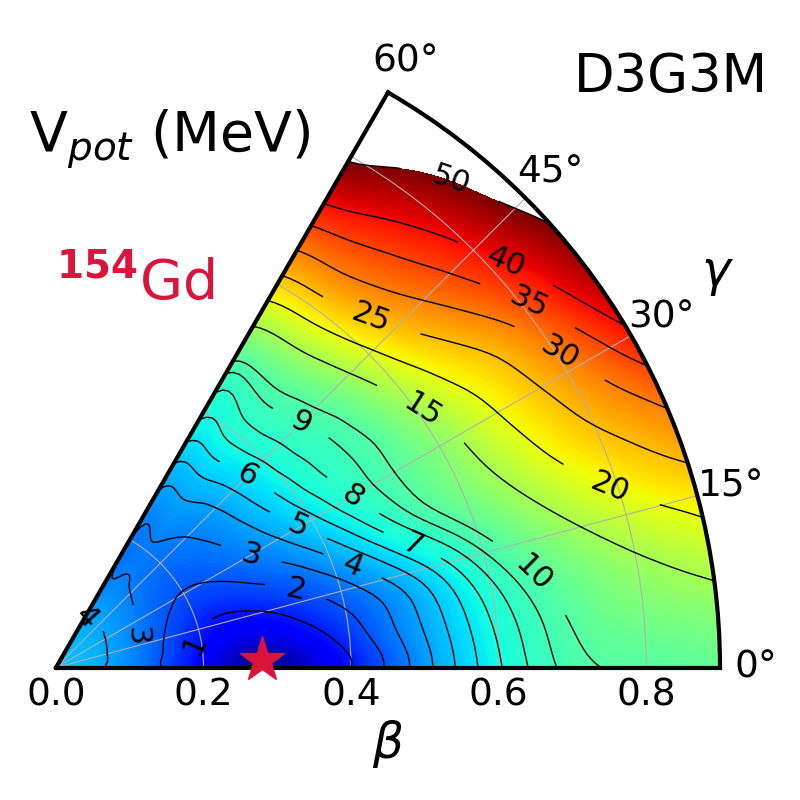}
     \caption{Potential energy surface $V_{pot}$, expressed in MeV, of the $^{154}$Gd obtained with HFB constrained calculations with the Gogny effective interaction D3G3M re-scaled to its energy minimum. The contour lines connect points of equal energy.}
     \label{fig:PES}
 \end{figure}

 In Fig.\ref{fig:PES}, we  show the potential energy surface of $^{154}$Gd obtained from constrained HFB calculations using the Gogny D3G3M interaction. 
 We notice that $^{154}$Gd  has a well pronounced prolate minimum at $\beta\approx0.3$ (indicated by a star). All selected Gogny interactions provide very similar results for this nucleus and the energy minimum is located at essentially the same position in the $\beta,\gamma$ plane : the inclusion of the zero point energy does not seem to change the position of the minimum.
 \begin{figure}
     \centering
         \centering
           \includegraphics[width=0.9\textwidth]{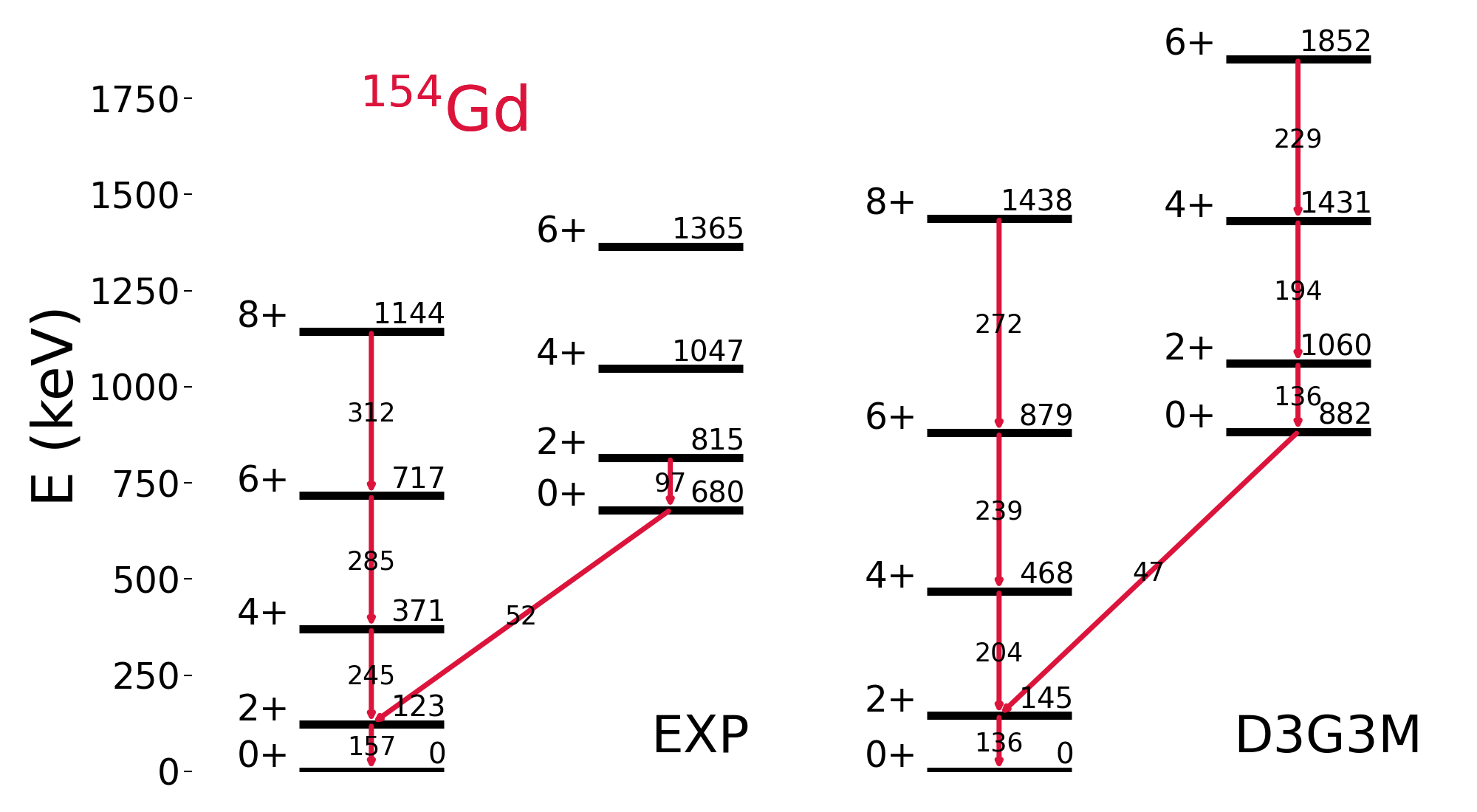}
         \caption{Experimental (EXP) data, taken from ENSDF database \cite{ENSDF}, and calculated (BH) level schemes (expressed in keV) with D3G3M and the corresponding electromagnetic transition strengths (in Weisskopf units) for $^{154}$Gd. }
     \label{fig:spectrum}
 \end{figure}

 We show in Fig.\ref{fig:spectrum} the low-energy spectrum of $^{154}$Gd (expressed in keV) obtained using the D3G3M interaction. On the same figure, we also report the experimental data taken from the ENSDF database \cite{ENSDF}. The method reproduces the overall structure quite well: both the ordering and the transitions, expressed in Weisskopf units, between states are in fair agreement.
 We observe that, in the higher energy part of the spectrum, the density of states is lower than the experimental value, even when a scaling factor of 1.3 is applied. This is a well-known issue related to the cranking approximation of vibrational masses \cite{Libert_1999}.

 \begin{figure}
     \centering
         \centering
           \includegraphics[width=0.49\textwidth]{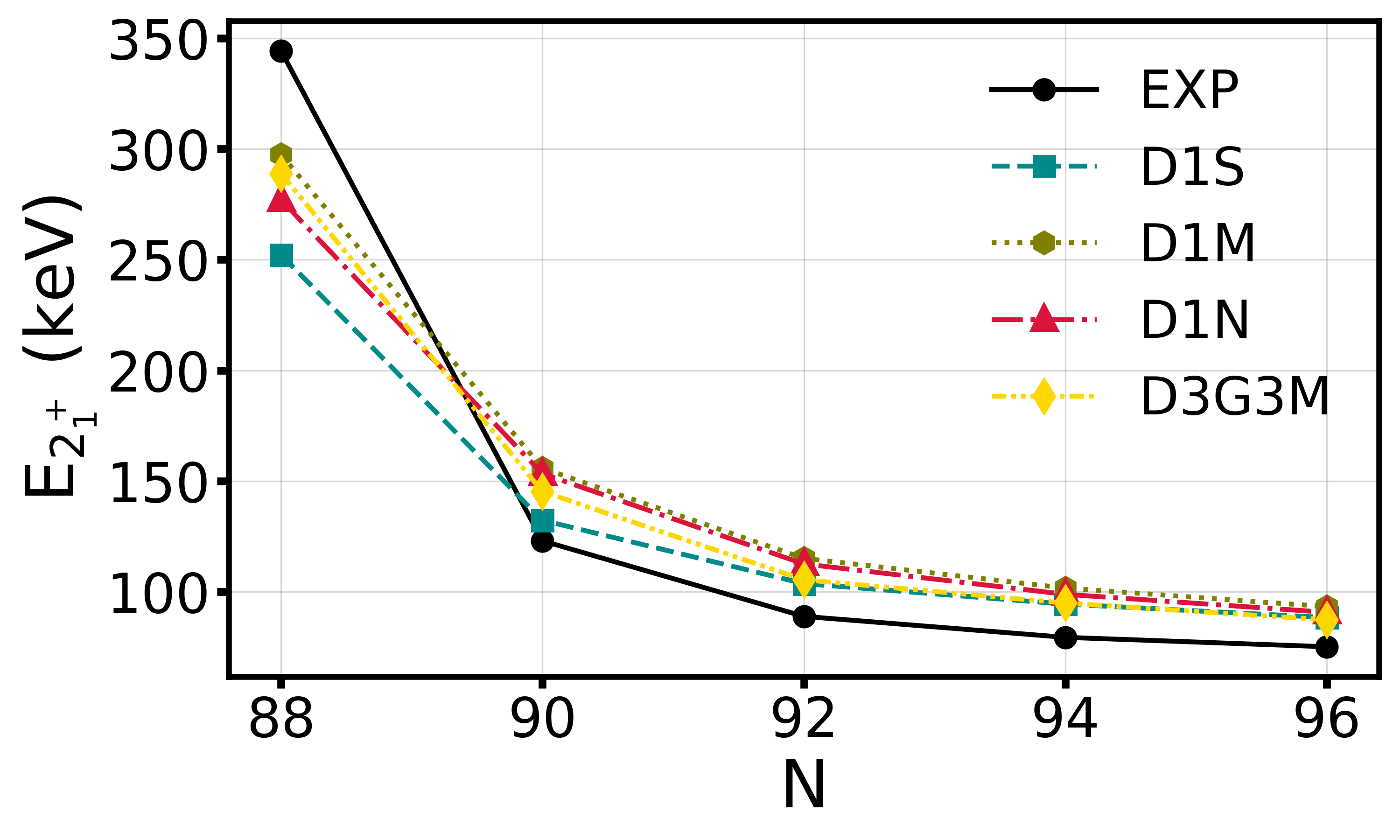}
           \includegraphics[width=0.49\textwidth]{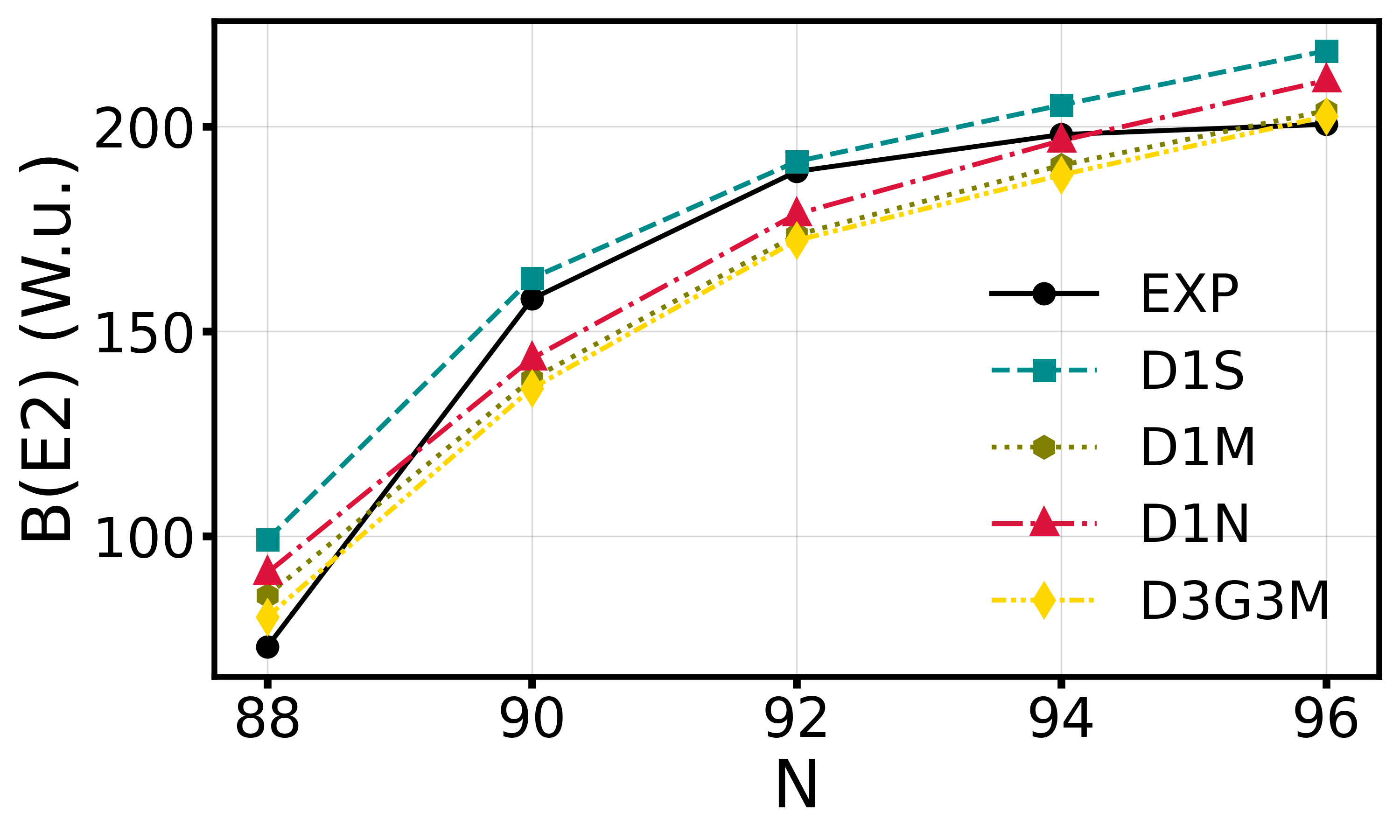}
         \caption{Evolution of the first excited $2^+$ level and the B$(E2;2^+ \to 0^+)$ transition strength for $^{152-160}$Gd. Dashed lines correspond to BH calculations  with various Gogny  interactions, while the solid line denotes the experimental data. }
     \label{fig:Gd_chain}
 \end{figure}

 Finally, in Fig.\ref{fig:Gd_chain}, we illustrate the evolution of the first $2^+$ excited state (left panel) and the corresponding  transitions, B$(E2;2^+ \to 0^+)$ (right panel) \cite{prochniak2009quadrupole}, along the even–even Gd isotopic chain, calculated with different effective Gogny interactions.
 We notice that although the results are very close to each other, the best agreements between theory and data (on average) is  obtained using Gogny D1S. A more detailed analysis is necessary in order to draw robust conclusions about the role of the underlying NN interaction.

 \section{Conclusions}

 To describe the collective excitations of even-even Gd nuclei and the associated electromagnetic transitions, we solved the Bohr Hamiltonian with quadrupole nuclear degrees of freedom only.
 The required input for the BH calculation, such as masses and potential  energy surfaces, has been obtained using constrained HFB calculations together with an effective Gogny interaction.
 We have performed a preliminary sensitivity test of the results using four selected Gogny interactions and we found that the results for Gd isotopes are quite robust, showing little dependence on the interaction, although a slight better agreement is observed for the Gogny D1S interaction.

 \section*{Acknowledgments}

 We thank J. Libert for helpful discussions on BH and S. P\'eru for her support in analysing the results.

 \bibliographystyle{ieeetr}
 \bibliography{biblio}

\end{document}